\documentclass[aps,prd,twocolumn,superscriptaddress,nofootinbib,amsmath,amssymb,floatfix,10pt]{revtex4-2}

\usepackage{booktabs}
\usepackage{graphicx}
\usepackage{xurl}
\usepackage[hidelinks]{hyperref}
\graphicspath{{figures/}{paper_draft/workflow_architecture_v1/figures/}}

\hypersetup{
  pdftitle={A Content-Addressed Workflow for Reproducible DANTE Gravitational-Wave Anomaly Analysis},
  pdfauthor={Luca Cirfeta},
  pdfsubject={Architecture and verification of the DANTE local scientific workflow},
  pdfkeywords={gravitational waves, anomaly detection, reproducibility, provenance, scientific workflows}
}

\newcommand{\dante}{\textsc{Dante}}

\begin{document}

\title{\texorpdfstring{A Content-Addressed Workflow for Reproducible\\
DANTE Gravitational-Wave Anomaly Analysis}{A Content-Addressed Workflow for
Reproducible DANTE Gravitational-Wave Anomaly Analysis}}

\author{Luca Cirfeta}
\thanks{ORCID: \href{https://orcid.org/0009-0000-1235-3186}{0009-0000-1235-3186}}
\affiliation{Independent Researcher, Rome, Italy}
\date{September 8, 2026}

\begin{abstract}
Unsupervised analysis of gravitational-wave detector data combines expensive
scientific stages with data acquisition, provenance checks, statistical
calibration, follow-up products, and reporting.  Reproducing such an analysis
requires more than preserving model weights or a final candidate table: the
execution graph, scientific contracts, retry semantics, and verification
boundary must also be explicit.  We present a local workflow layer for the
Domain-Adaptive Network for Transient Evaluation (\dante{}) that represents a
corrected O4a analysis as a content-addressed, 15-stage directed acyclic graph.
The architecture separates immutable stage evidence from mutable operational
progress, binds command-line and browser interfaces to the same run identity,
and resumes only work compatible with the frozen contract.  The tagged v1
release records all 15 stages as verified in a machine-readable receipt and
passed a documented human usability gate.  The release verifies adopted
scientific artifacts rather than recomputing them, and therefore makes no new
claim of global significance, astrophysical discovery, or public real-time
operation.  This paper describes the architecture, its fail-closed controls,
and the bounded evidence supporting reproducible local use.
\end{abstract}

\maketitle

\section{Introduction}
\label{sec:introduction}

Computational reproducibility in detector characterization depends on the
identity of data, code, configuration, execution environment, and intermediate
artifacts.  A scientifically correct stage can still be difficult to reproduce
when its ordering, retry behavior, or upstream evidence is implicit.  This work
addresses that operational layer for \dante{} without changing what the
underlying anomaly analysis measures or how its statistical decisions are
validated.

The contribution has four parts.  First, a versioned workflow contract makes
stage order, dependencies, expected artifacts, verifier commands, outcome
visibility, and resumability explicit.  Second, a durable state model separates
an immutable history of attempts and artifact hashes from replaceable progress
information.  Third, command-line and local-web interfaces resolve the same
orchestrator and therefore resolve the same run identity.  Fourth, a final receipt
is issued only from a complete, currently verified artifact graph.  The
implementation is frozen by the annotated tag
\texttt{dante-workflow-productization-v1}; its machine-readable release receipt
is distributed in the tagged source tree.

\section{Related work and design position}
\label{sec:related}

General workflow systems already express computations as dependency graphs.
Snakemake derives a directed acyclic graph from file-oriented rules and uses it
to schedule or resume work across local and cluster environments
\cite{koster2012snakemake}; Nextflow combines a dataflow model with portable,
container-oriented execution across computational platforms
\cite{ditommaso2017nextflow}.  Current Snakemake records job provenance and, by
default, considers changes in code, inputs, parameters, software environment,
and modification time when deciding whether to rerun a job
\cite{snakemake2026cli}.  Nextflow computes per-task hashes from execution
metadata that include the task inputs, script, and software environment, and
\texttt{-resume} requires both a matching cache entry and retained outputs
\cite{nextflow2026cache}.  Table~\ref{tab:comparison} therefore compares
operational semantics rather than claiming that content-aware reuse is new.

\begin{table*}[t]
\caption{Design-scope comparison.  Entries describe the documented core
semantics relevant to this work, not a feature ranking of the three systems.}
\label{tab:comparison}
\begin{ruledtabular}
\begin{tabular}{p{0.12\textwidth}p{0.22\textwidth}p{0.22\textwidth}p{0.28\textwidth}}
Dimension & Snakemake & Nextflow & This \dante{} workflow layer \\
\hline
Reuse identity & Per-job provenance/rerun triggers over code, inputs,
parameters, environment, and time/checksum metadata & Per-task hash over
session and task metadata, inputs, script, container or package environment &
One workflow-wide canonical run key over the frozen workflow contract, source
identity, roots, and concrete stage-command digests; stage artifacts have
separate receipts \\
Retry semantics & Re-evaluate the DAG and rerun jobs selected by provenance
triggers & \texttt{-resume} reuses a task only
when cache metadata and required work-directory outputs remain valid & Preserve
every terminal/incomplete attempt; retry under the same run key creates a fresh
attempt identity and re-verifies recorded artifact bytes \\
Concurrent ownership & General scheduler/executor model for jobs & Concurrent
dataflow execution; the default per-session cache store admits one
reader/writer & One durable local worker lease per run key; stale active work is
recorded as interrupted before replacement \\
Operator surface & General-purpose command-line workflow interface and plugin
ecosystem & General-purpose command-line interface and execution history &
Local CLI and loopback browser invoke the same domain controller and resolve
the same run key and evidence graph \\
Release boundary & User-defined workflow targets and output validation &
User-defined publish and reporting behavior & A report is refused until every
declared scientific-stage verifier and current-byte artifact check passes \\
\end{tabular}
\end{ruledtabular}
\end{table*}

The workflow described here is not a replacement for either general-purpose
engine.  It is a domain-specific control and evidence layer around already
frozen \dante{} commands and verifiers.  Its contribution is the tested
composition of workflow-wide identity, append-only attempt evidence, a durable
single-owner local lease, one CLI/browser control path, and a fail-closed
scientific-report boundary.  It does not claim algorithmic novelty for hashing,
caching, DAG scheduling, or content addressing.

The content-addressed design itself has direct prior art in software versioning
and build systems.  Git exposes a content-addressable object database
\cite{chacon2014progit}; Bazel separates an action cache from a
content-addressable store of output files \cite{bazel2026cache}; and Nix uses a
functional deployment model with immutable outputs and input-derived store
identities \cite{dolstra2006nix}.  The present work adapts these ideas to a
bounded scientific workflow in which cached evidence is insufficient until its
current bytes and domain-specific verifier are checked again.

The W3C PROV data model supplies a general vocabulary for entities, activities,
agents, derivations, and provenance bundles \cite{moreau2013provdm}.  The
\dante{} receipts instantiate a narrower application-specific JSON model and
do not claim PROV conformance or interchangeability.  Likewise, the FAIR
principles emphasize findable, accessible, interoperable, and reusable digital
research objects, including data, algorithms, tools, and workflows
\cite{wilkinson2016fair}.  Content-addressed receipts and documented reuse
support traceability within this release, but they do not by themselves
establish FAIR compliance, archival persistence, or repository-level
discoverability.

Public strain and data-quality products distributed through the
Gravitational-Wave Open Science Center make independent detector-data analyses
possible \cite{abbott2021gwosc}.  The O4a-specific release supplies H1 and L1
strain, segment, checksum, and auxiliary-data products and identifies the
released calibrated-strain channels \cite{gwosc2025o4a}.  The bounded public
smoke uses that public-data boundary to test installation and interface parity.
Full corrected-O4a reproduction remains a stronger requirement: it additionally
depends on the frozen archive, scientific runtime, contracts, and intermediate
evidence named by this workflow.

\section{Scope and threat model}
\label{sec:scope}

The protected failure modes are accidental parameter or source drift,
incompatible resume, duplicate workers for one identity, incomplete attempts,
changed verified artifacts, interface-specific identities, and premature
presentation of outcome-bearing files.  The trusted computing base still
includes the local operating system, file system, Python environment, frozen
scientific commands, and their stage verifiers.  The design does not claim
tolerance to arbitrary hardware failure, distributed execution, hostile
multi-user access, or equivalence between CPU and CUDA environments.  It is a
local single-user controller, not an alerting service.

The workflow also distinguishes reproducibility of an execution contract from
repeatability of a numerical result on arbitrary hardware.  A device choice
represented by the frozen command or configuration, a repository-state
change, or a root-path change contributes to a different run identity.  Such
identities may be compared by a separately defined scientific validation, but
the orchestration layer never assumes their equivalence.

\section{Scientific and orchestration boundaries}
\label{sec:boundaries}

Scientific configuration and stage verifiers remain authoritative.  The
orchestration layer orders those stages, records their identities, and refuses
incompatible evidence; it does not duplicate scoring, threshold, bootstrap,
coincidence, or environmental-monitor logic.  The released workflow adopts and
re-verifies existing corrected-O4a artifacts.  It does not independently
recompute the scientific chain.  In the release receipt every stage therefore
has execution mode \texttt{ADOPTED\_VERIFIED\_EXISTING}: calculation commands
were not rerun, while the existing verifier associated with each stage was
replayed.  This distinction is part of the receipt rather than an editorial
qualification added after execution.

\noindent\begin{minipage}{\columnwidth}
The adopted current catalogue is the corrected-O4a native-classification run:
\par\smallskip
{\raggedright\noindent\footnotesize
\nolinkurl{248df45d52faac527cb0e33932010b9fa8a354b2ace73bcdd42d714e49f4a1ef}\par}
\smallskip
It contains 10,942 detector--GPS candidates, comprising 5,406
\texttt{ROBUST}, 2,344
\texttt{AMBIGUOUS}, and 3,192 \texttt{BACKGROUND}.  The 10,429-candidate v6
detector-aware catalogue (6,365/1,275/2,789) remains immutable as the
\emph{historical comparison baseline}; it is consumed by the final-comparison
stage and is not the adopted current classification snapshot.  The release
receipt names both roles and binds the corrected classification and comparison
run directories separately.  Thus \texttt{PASS\_VERIFIED\_WORKFLOW} certifies
the current 10,942-candidate graph while preserving, rather than promoting, the
10,429-row baseline.
\end{minipage}

\section{Workflow architecture}
\label{sec:architecture}

\begin{figure*}[t]
\centering
\includegraphics[width=\textwidth]{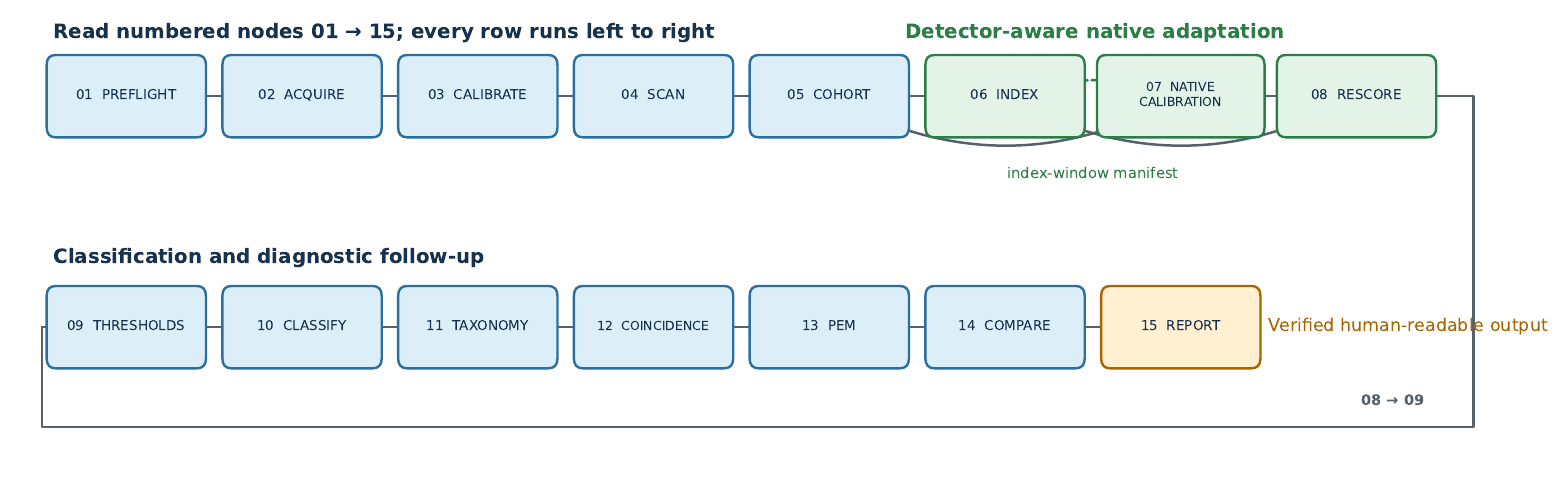}
\caption{The frozen 15-stage workflow, numbered in execution order with both
rows read left to right.  Solid arrows are verified stage dependencies.  The
dashed green edge is the content-digested index-window manifest consumed by
native calibration; it makes the exclusion relationship explicit before both
native products enter rescoring.}
\label{fig:dag}
\end{figure*}

The frozen graph contains 15 stages from preflight and acquisition through
calibration, scanning, native adaptation, follow-up, comparison, and reporting.
The native-calibration node depends both on the cohort and on the window
manifest consumed by the native index, preserving the frozen exclusion guard
before rescoring.  Figure~\ref{fig:dag} is generated directly from the
versioned workflow configuration; the layout is fixed for readability, while
the stage and dependency sets are validated against the contract at generation
time.

The graph can be read in four blocks.  \texttt{PREFLIGHT} and
\texttt{ACQUIRE} establish infrastructure and data-manifest evidence.
\texttt{CALIBRATE} and \texttt{SCAN} bind the primary thresholds and candidate
catalogue.  \texttt{COHORT} then feeds \texttt{INDEX};
\texttt{NATIVE\_CALIBRATION} requires both the cohort and the
content-digested index-window manifest before \texttt{RESCORE}.  The remaining
linear chain---\texttt{THRESHOLDS}, \texttt{CLASSIFY}, \texttt{TAXONOMY},
\texttt{COINCIDENCE}, \texttt{PEM}, \texttt{COMPARE}, and
\texttt{REPORT}---preserves the existing scientific stage boundaries through
the human-facing report.  A downstream stage cannot start until each declared
\texttt{VERIFIED\_STAGE} dependency is currently verified.

\section{Identity, state, and provenance}
\label{sec:provenance}

The workflow specification is canonicalized as sorted, compact JSON and
self-digested with SHA-256.  Loading it verifies that digest, the DAG, required
policy fields, and the path and SHA-256 identity of every referenced scientific
configuration.  The run contract then binds the workflow identifier, workflow
contract digest, repository source identity, repository/raw/cache roots, and
the digests of the concrete run and verifier commands.  Its canonical SHA-256
is the run key,
\begin{equation}
  K_{\mathrm{run}} = H\!\left(\operatorname{canonicalJSON}
  (C_{\mathrm{run}})\right),
\end{equation}
and the durable directory is named \texttt{workflow\_}$K_{\mathrm{run}}$.
Consequently, resume is an operation within one frozen identity; a material
contract, source, environment, command, or path change selects another
identity instead of silently reusing prior evidence.

One durable worker lease owns a run key at a time.  The lease records process
and host identity, start time, and a random token; the browser never owns it.
Each stage execution receives a fresh attempt identifier and an exclusive
attempt directory.  Start, artifact, terminal, and interruption events are
appended to a JSON-lines ledger.  Recorded artifacts carry logical name,
resolved path, and SHA-256 digest.  A stage reaches \texttt{VERIFIED} only after
its declared outputs are present, its verifier exits successfully, and the
terminal event is durably appended.

Reuse is verification, not trust in a cached status bit.  Before a verified
stage is skipped, and before its evidence satisfies a dependency, the recorded
artifact bytes are hashed again.  A changed artifact, reused attempt identity,
missing declared output, conflicting contract, or attempt path that would
overwrite preserved evidence is rejected.  If a worker disappears, acquiring
a replacement lease records its active attempt as \texttt{INTERRUPTED}; retry
keeps the run key but creates a new attempt directory.  Mutable progress JSON
may be atomically replaced because it is explicitly operational state.  It is
not accepted as scientific evidence.

\begin{figure*}[t]
\centering
\includegraphics[width=0.96\textwidth]{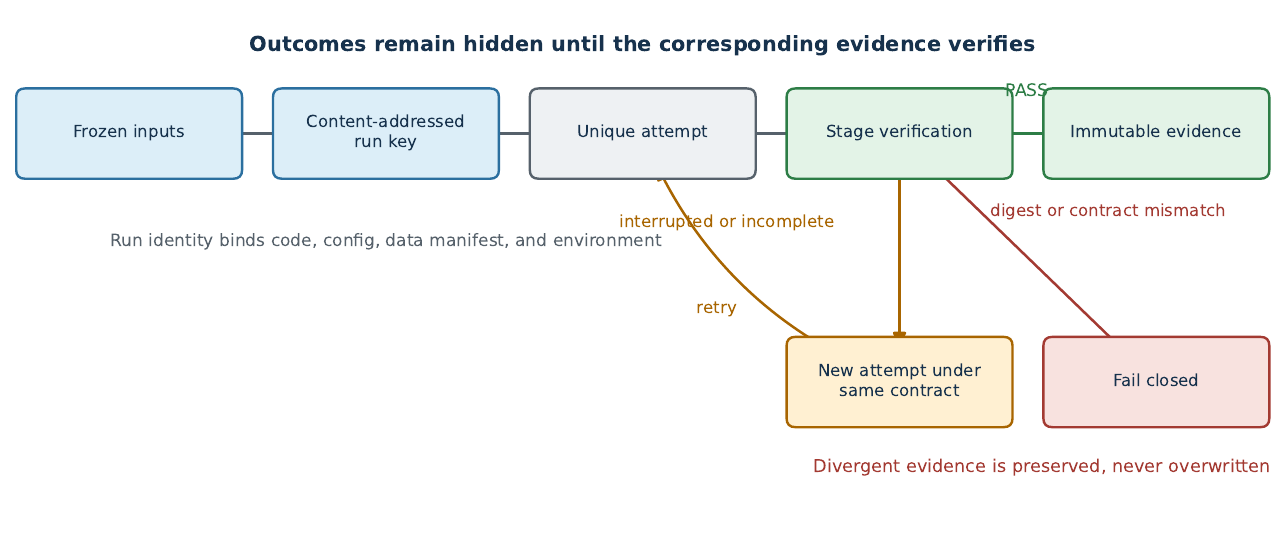}
\caption{Evidence lifecycle.  Incomplete work receives a new attempt under the
same contract, whereas a digest or contract mismatch fails closed.  Verified
evidence is immutable and outcome-bearing fields remain hidden until the
corresponding gate passes.}
\label{fig:evidence}
\end{figure*}

\section{Operator interfaces and recovery}
\label{sec:interfaces}

The command-line and browser interfaces invoke the same orchestrator.  The
browser is a loopback-only reader and controller; the workflow worker remains a
separate process launched with sealed standard-output and standard-error files.
The HTTP handler performs only bounded control and read operations.  Closing
the browser or UI server therefore leaves the worker lease and scientific
process intact; a new UI instance reconstructs state from the ledger.

The command line exposes plan, preflight, run, resume, status, verification,
report generation, bounded execution through a named stage, and explicit
single-stage repair.  Repair does not bypass dependencies.  Cooperative stop
is checked between atomic run-plus-verification stages, so an incomplete stage
does not produce a final report.  The UI maps the same operations to guided
controls and shows immutable scientific configuration digests rather than
editable scientific parameters.

For the bounded public smoke, the UI probes the worker environment and
recommends CUDA when a compatible NVIDIA device is available; CPU remains the
portable documented path.  The two choices have distinct run identities and
no numerical-equivalence claim is made.  The guided page reports current
phase, phase-based completion percentage, an explicitly approximate ETA,
worker state, live output, detailed logs, and recovery instructions.  Once the
receipt verifies, a dedicated results view separates the human-readable report
from the machine-readable technical receipt and explains the role of each.

\begin{figure*}[t]
\centering
\includegraphics[width=0.96\textwidth]{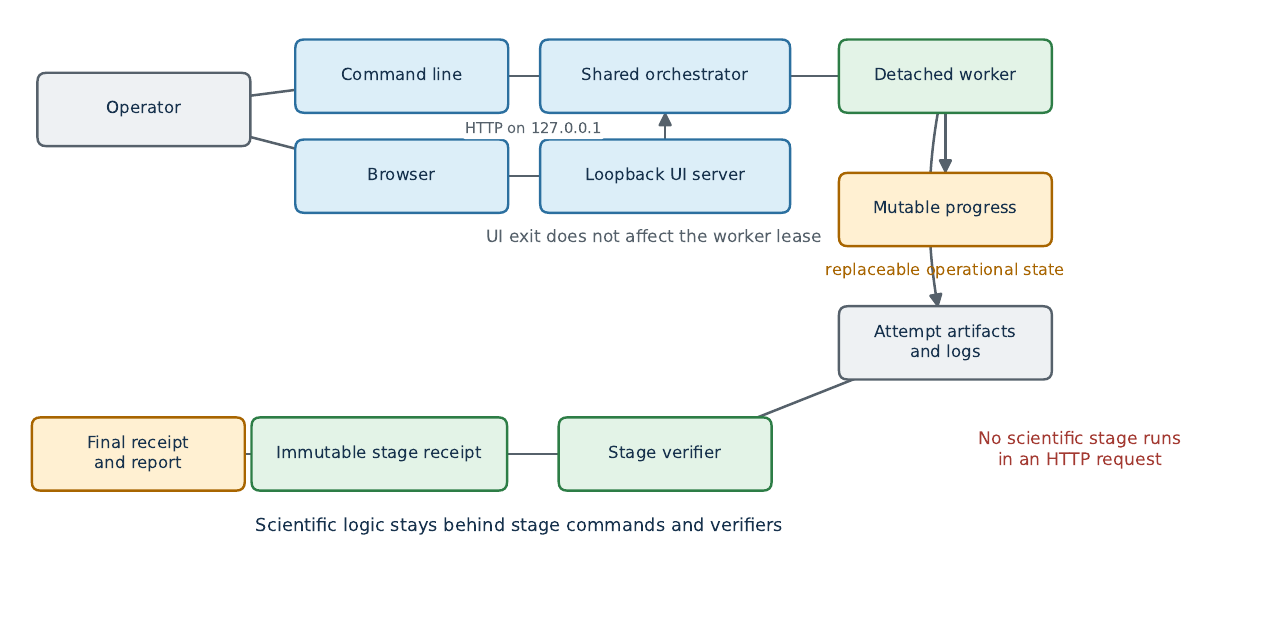}
\caption{Runtime boundary shared by the command-line and browser paths.  The
loopback UI is a disposable controller; the detached worker, durable attempts,
and stage verifiers remain authoritative.  Scientific stage logic is not
executed inside HTTP request handlers.}
\label{fig:runtime}
\end{figure*}

\section{Verification methodology}
\label{sec:verification}

Verification evidence includes schema and process tests, clean-clone public
smoke execution, command-line/browser identity parity, failure injection,
interruption recovery, final receipt verification, and human usability
acceptance.  Each test population and limitation will be reported separately;
technical smoke evidence will not be represented as a full scientific rerun.

The recovery checkpoint injected five stage-runner failures---network timeout,
CUDA error, dependency error, disk exhaustion, and verifier failure---and
checked that retry retained the run identity and prior attempts.  Additional
tests changed already verified receipts, terminated an inert owned child,
failed lease synchronization, and injected failures while appending terminal
and orphan-recovery events.  These are controlled engineering faults; they are
not observations of failures during the scientific O4a calculation.

The clean-clone checkpoint used an isolated Python 3.11 environment and a
two-window public H1/L1 smoke.  The packaged command-line and Waitress paths
produced byte-identical receipts.  Their recorded maximum absolute score
difference was zero under an existing absolute tolerance of $2\times10^{-7}$,
with no disposition mismatch.  This bounded replay checks installation,
acquisition, existing scoring, interface parity, and retry behavior; it does
not cover the full raw archive or the complete corrected-O4a graph.

Finally, release verification replays each stage verifier, reconstructs the
artifact graph, checks the index-consumption manifest, and derives the report
only after all stages pass.  The usability gate was deliberately human: an
initial attempt failed because actions, device choice, progress, logs, recovery,
and report/receipt roles were unclear.  After those defects were corrected, a
second WSL attempt accepted the guided workflow and requested only a separate
verification-results tab; that revision was implemented and reverified before
the gate was closed.

\section{Results}
\label{sec:results}

The tagged release receipt reports \texttt{PASS\_VERIFIED\_WORKFLOW} for the
15-stage graph.  All 15 verifier exit statuses are zero, no stage is incomplete,
and the index-consumption manifest is an exact match.  Table~\ref{tab:identity}
records the content identities.  The runtime receipt and canonical Git blob
were recorded as byte-identical; the receipt self-digest is a field inside that
object, whereas the canonical-byte SHA-256 identifies its serialization.

\begin{table*}[t]
\caption{Frozen productization-v1 release identity.  Values are transcribed
from the versioned release result and machine-readable receipt.}
\label{tab:identity}
\begin{ruledtabular}
\begin{tabular}{ll}
Field & Value \\
\hline
Workflow & \texttt{dante-o4a-corrected-productization-v1} \\
Run key & \texttt{3c9ad0765b4d1f9f9a49cf105ca223058aafdba42ac25793d0ecff1d1d384e94} \\
Contract digest & \texttt{1d7270c0bcd15b3c344f94bb37fe82033b67bf4ad12615bb50b5efab43422cdd} \\
Artifact-graph digest & \texttt{188a93c5b73cac50987e208f289729f4b17650e7f82fd0d91ecb1b7a736049fa} \\
Receipt self-digest & \texttt{3b062e473e70e348ba5f39cbca17676cf63eaa575ea1d95639280bd664440aa4} \\
Canonical receipt bytes SHA-256 & \texttt{a2061ffbd96a25421968bfc0fd14798a72e31152e50e4481aa7f1c6a825d9d86} \\
Derived-report SHA-256 & \texttt{3711b278b974cf0da2206b0136d5b26250e61c6492e2a562f66dbabf2298d9e2} \\
\end{tabular}
\end{ruledtabular}
\end{table*}

The acceptance evidence is intentionally layered rather than summed as one
test count.  Table~\ref{tab:tests} reconciles the counts by source revision and
platform.  The increase from 104 to 121 collected workflow tests reflects later
UI, recovery, and verification-result coverage; it is not an additional sample
of independent statistical trials.  The sole WSL skip is a Windows-specific
process-lifecycle assertion.  The 121-test release corpus therefore appears as
121 passes on Windows and as 120 passes plus one platform skip on WSL.

\begin{table*}[t]
\caption{Workflow regression checkpoints.  Evidence paths are repository
relative and versioned; the counts within or across rows must not be summed.}
\label{tab:tests}
\begin{ruledtabular}
\begin{tabular}{p{0.13\textwidth}p{0.10\textwidth}p{0.09\textwidth}p{0.12\textwidth}p{0.40\textwidth}}
Checkpoint & Source revision & Windows & Fresh WSL & Versioned evidence \\
\hline
Packaged CLI/UI parity & \texttt{143b7539} & 104 pass & 103 pass, 1 skip &
Parity checkpoint JSON \\
Human-gate release & \texttt{e8f2098e} & 121 pass & 120 pass, 1 skip &
Release result note \\
\end{tabular}
\end{ruledtabular}
\end{table*}

The exact evidence files are
\path{artifacts/dante_workflow/public_smoke_ui_checkpoint_2026-09-06.json} and
\path{docs/DANTE_WORKFLOW_PRODUCTIZATION_RESULT.md}.  The packaged parity
checkpoint also records a 5936-byte technical receipt with
SHA-256
\nolinkurl{e91c744891dc70ff2aa20e78c11b2f1f5ae4342a445d6a6a9bba90a479308e58}
from both interfaces.  The release tests are versioned local acceptance
evidence; this paper does not substitute a CI badge for those recorded
platform-specific executions.

The result is therefore an engineering claim: the tagged local workflow can
identify, inspect, resume, verify, and report the adopted artifact graph under
the recorded contract.  It does not interpret candidate outcomes or convert
the successful verifier matrix into detection significance.

\section{Limitations}
\label{sec:limitations}

The release is local and single-user, relies on pinned external scientific
artifacts, and distinguishes rather than equates CPU and CUDA identities.  Its
adopted-stage verification does not substitute for a new end-to-end O4a
recalculation.  Coincidence and environmental-monitor products remain
diagnostic follow-up evidence, and no global-significance or discovery claim
follows from workflow verification.

Content addressing detects known identity changes but cannot prove that an
external program is scientifically correct, that a hardware or library defect
is absent, or that all semantically relevant environmental state has been
captured.  Absolute paths are intentionally part of the present run contract;
this prevents ambiguous reuse but makes relocation a new identity.  The worker
model is local rather than distributed, the UI has no remote-authentication
model, and recovery is stage-granular except where an underlying scientific
stage implements its own safe checkpointing.

The canonical receipt-byte digest in Table~\ref{tab:identity} refers to the Git
blob and runtime serialization.  A checkout configured to translate text line
endings can materialize different working-tree bytes while retaining the same
parsed JSON and self-digest.  Byte-level comparison should therefore use the
canonical blob or an LF-normalized checkout; independent receipt verification
checks the canonical object content.

The public smoke is deliberately small and uses existing scoring behavior.  It
does not validate full-archive acquisition, reproduce the historical CUDA
software stack exactly, or establish CPU/CUDA equivalence.  Likewise, verifier
replay authenticates current bytes and stage-specific acceptance conditions;
it is not an independent recalculation of every adopted value.

\section{Reproducibility}
\label{sec:reproducibility}

The software baseline is the annotated tag
\texttt{dante-workflow-productization-v1}.  The portable public path uses
Python 3.11, \texttt{requirements-cpu.txt}, and a clean WSL-native checkout;
\texttt{requirements-ui.txt} adds the optional browser interface.  From the
repository root, the bounded command-line sequence is
{\small
\begin{verbatim}
runner=scripts/run_dante_workflow_clean_clone.py
python $runner --mode plan --device cpu
python $runner --mode local --device cpu
python $runner --mode verify --device cpu
\end{verbatim}
}
and the same smoke can be controlled at \texttt{127.0.0.1:8765} with
{\small
\begin{verbatim}
ui=scripts/run_dante_workflow_ui.py
python $ui --public-smoke
\end{verbatim}
}
after installing the optional requirements.  Repeating local mode checks and
reuses verified engine attempts; incomplete attempts remain preserved and are
replaced by a new attempt.  The expected products are a technical receipt,
paired-replay evidence, and per-attempt logs under a content-addressed
directory.

The full release receipt can be checked independently with
{\small
\begin{verbatim}
python scripts/verify_dante_workflow.py \
  --release artifacts/dante_workflow/\
productization_v1_release.json
\end{verbatim}
}
This verifies the serialized release object.  Re-executing the full 15-stage
corrected-O4a workflow additionally requires the frozen raw archive, canonical
scientific runtime, and paths named by the workflow configuration; the public
two-window smoke is not a substitute.  The final arXiv bundle will include a
source manifest and cryptographic digests but no automated publication step.
The tagged source, including the named scripts, tests, configuration, and
versioned evidence, is available at
\url{https://github.com/lucacirfeta/dante-gravi-signal-ml/tree/dante-workflow-productization-v1}.

\section{Conclusion}
\label{sec:conclusion}

The implemented workflow makes the operational contract of a complex anomaly
analysis explicit and machine-verifiable while preserving the authority and
limits of its scientific stages.  Content-derived run identity, append-only
attempt evidence, current-byte verification, detached operation, and one
shared CLI/UI control path address concrete reproducibility and usability
failures without redefining the analysis.  The tagged release verifies a
15-stage adopted artifact graph and a bounded clean-clone smoke.  It neither
recomputes the full corrected-O4a chain nor supplies a claim of global
significance, astrophysical discovery, or public real-time operation.

\bibliographystyle{apsrev4-2}
\bibliography{references}

\end{document}